\documentstyle[12pt,epsfig]{article}
\newcommand{\beq}{\begin{equation}}
\newcommand{\eeq}{\end{equation}}
\newcommand{\bea}{\begin{eqnarray}}
\newcommand{\eea}{\end{eqnarray}}
\def\lsim{\raise0.3ex\hbox{$\;<$\kern-0.75em\raise-1.1ex\hbox{$\sim\;$}}}
\def\gsim{\raise0.3ex\hbox{$\;>$\kern-0.75em\raise-1.1ex\hbox{$\sim\;$}}}
 
\begin{document}
\begin{center}
{\large{\bf Neutrino Spectra from Low and High Luminosity Populations of
Gamma Ray Bursts}}\\ 
\medskip
{Nayantara Gupta \footnote{Email: nayan@physics.unlv.edu, 
Phone No. 1 702 8951702, Fax No. 1 702 8950804 }, 
Bing Zhang \footnote{Email: bzhang@physics.unlv.edu}}\\
{Department of Physics, 
University of Nevada Las Vegas, Las Vegas, NV 89154, USA}
\end{center}
\begin{abstract}
The detection of  GRB 060218 at z=0.033 by {\em Swift} within 1.5 years
of operation, together with the detection of GRB 980425 at $z=0.0085$ by
BeppoSAX, suggest that these low-luminosity (LL) GRBs have a much
higher event rate than the canonical high-luminosity (HL) GRBs, and
they form a distinct new component in the GRB luminosity function. 
We explore the contribution of this previously neglected GRB
population to the diffuse neutrino background within the internal
shock model and compare it with that of the canonical HL
population. By considering a wide range of distributions of various
parameters (e.g. luminosity, spectral break energy, duration, variability time,
Lorentz factor, redshift) for both populations, 
we find that although it is difficult to detect
neutrinos from the individual LL GRBs, the contribution of the LL
population to the diffuse neutrino background is more than the HL population
 above about $10^{8}$ GeV. 
\end{abstract}


\section{Introduction}
The possibility of high energy neutrino emission from GRBs in the
burst and the afterglow phases has been studied earlier by different
groups \cite{wax1,all,razzaque,raz2}.  It has been found that only the
nearby high luminosity (HL) GRBs (e.g. GRB 030329) are expected to be
detected individually by neutrino telescopes like ICECUBE
\cite{razzaque,ice1}. In general the diffuse neutrino background from
the whole GRB population is of great observational interests. Recently
a low-luminosity (LL), long-duration event, GRB 060218/SN 2006aj was
detected by {\it Swift} \cite{campana} at a redshift of 0.0331. 
Earlier, another low luminosity event,
GRB 980425/SN 1998bw was detected by {\it BeppoSAX} at an even lower
redshift of 0.0085 \cite{galama,kul}. 
These two LL events 
share some common characteristics such as low luminosity, long
duration, and low isotropic $\gamma$-ray energy. 
More importantly, their detections at very low redshifts within a
relatively short period of time (1.5 years for {\em Swift} and 6 years
for {\em BeppoSAX}) imply that they have a much higher
event rate than the canonical HL GRBs \cite{cobb,pian,soderberg,liang}.
A straightforward estimate of 
the local rate of LL-GRBs $\rho_{0}^{\rm LL}$ can be derived from
\beq
\rho_{0}^{\rm LL}V(z<0.033)\Big(\frac{\Omega^{Beppo}}{4\pi}T^{Beppo}
+\frac{\Omega^{Swift}}{4\pi}T^{Swift}\Big)\sim 2, 
\eeq
where $V(z<0.033)\sim1.2\times10^{-2}{\rm Gpc}^3$ is the volume
enclosed by $z=0.033$, $\Omega^{Beppo}=0.123$ and $\Omega^{Swift}=1.33$
are the solid angles of the GRBM on board {\it BeppoSAX} and the BAT
on board {\it Swift} respectively. The durations 
of observation are $T^{Beppo} \sim 6$ yr and $T^{Swift}\sim 1.5$ yr (when
this paper has been written) for {\it BeppoSAX} and {\it Swift},
respectively. The number 2 on the right hand side of Eq.(1) accounts
for the detection of GRB 980425 and GRB 060218. This gives a rough
estimate of $\rho_{0}^{\rm LL}$ as $800 {\rm Gpc^{-3}yr^{-1}}$. Several
groups independently estimated this event rate\cite{cobb,soderberg}, 
e.g. $\sim (300-1000) {\rm Gpc^{-3} yr^{-1}}$, which are
comparable to the rate of LL GRBs
$550_{-430}^{+700} {\rm Gpc^{-3} yr^{-1}}$ obtained by Liang et
al.\cite{liang}. These are all much higher than the event rate of HL
GRBs, $\rho_{0}^{\rm HL}\sim 1 {\rm Gpc^{-3}yr^{-1}}$\cite{schmidt}, and
its simple extrapolation to the LL regime assuming a same population,
e.g. $\rho_{0} \sim 10 {\rm Gpc^{-3}yr^{-1}}$\cite{guetta}.
If one assumes that the two LL GRBs within $z \le 0.033$
belong to the population of HL GRBs, then the expected number of
detection of GRBs within $z \le 0.033$ in the {\it BeppoSAX} and {\it
Swift} era is nearly 0.0026 for $\rho_{0}^{\rm HL}\sim 1 {\rm
Gpc^{-3}yr^{-1}}$, and is 0.026 for $\rho_{0} \sim 10 {\rm
Gpc^{-3}yr^{-1}}$. The Poisson probability of detecting two events
during the observational period is $3.4 \times 10^{-6}$ for the former
and $3.3\times 10^{-4}$ for the latter. Hence, the detection of GRB
060218 strongly suggests the presence of a
distinct population (LL), which is not a subset of the canonical HL
population.  This implies two distinct components in the GRB luminosity
function\cite{liang}. One may characterize the luminosity function of
each population by a broken power law, i.e.
\beq
S(L)=S_{0}\Big[\Big(\frac{L}{L_b}\Big)^{\alpha_1}+
\Big(\frac{L}{L_b}\Big)^{\alpha_2}\Big]^{-1}
\eeq  
The spectral indices and break luminosities of this function for LL
and HL populations have been derived/constrained in \cite{liang}. The
values of $\alpha_1$, $\alpha_2$, $L_b$ are $0.6$, $4.5$,
$10^{47}$erg/sec and $1.15$, $2.5$, $10^{51}$erg/sec for LL and HL
GRBs respectively. The values of $S_{0}$ depend on the upper and lower
limits in the values of luminosity for each population.
 
Although the high energy neutrino flux of these individual LL GRBs is
low, the contribution to the diffuse neutrino flux by the whole LL
population is non-negligible due to their very high event rate. In
this paper we explore the neutrino emission of this hitherto ignored
LL-GRB population, and compare the contributions of the LL- and HL-
GRB populations to the diffuse neutrino background. 

\section{Diffuse Neutrino Spectra from HL and LL GRBs} 
We consider the prompt high energy neutrino emission from GRBs in
the burst phase due to internal shocks. 
Protons and photons interact to produce pions which
subsequently decay to neutrinos, $p+\gamma\rightarrow
\pi^{+}\rightarrow \mu^{+}+\nu_{\mu}\rightarrow
e^{+}+\nu_e+\bar\nu_{\mu}+\nu_{\mu}$. This is the dominant mechanism
of high energy neutrino production inside GRBs. We derive the neutrino
spectrum using the power law nature of photon spectrum as observed by
different satellite experiments BATSE \cite{bat}, HETE \cite{hete},
Swift \cite{swift}. GRB 060218 is an X-ray flash (XRF)\cite{campana}. 
Sakamoto et al.\cite{saka} have discussed in detail about the
characteristics of 45 GRBs observed by HETE-2 (High Energy Transient
Explorer)\cite{hete}, some of them were XRFs. They found no
statistically significant deviation in the distribution of photon
spectral indices depending on the type of the source. We therefore
describe the comoving-frame photon spectrum for both HL and LL GRBs as
\beq
\frac{dn_{\gamma}}{d\epsilon_{\gamma}} 
=A \left\{ \begin{array}{l@{\quad \quad}l}
{\epsilon_{\gamma}}^{-\gamma_1} &
\epsilon_{\gamma}\leq \epsilon_{br,c}\\{\epsilon_{br,c}}^{\gamma_2-\gamma_1}
{\epsilon_{\gamma}}^{-\gamma_2} & \epsilon_{\gamma}>\epsilon_{br,c}
\end{array}\right.
\eeq 
The break energy of the photon spectrum in the comoving frame
$\epsilon_{br,c}$ is related to its value in the source rest frame as
$\epsilon_{br}=\Gamma\epsilon_{br,c}$, where $\Gamma$ is the bulk
Lorentz factor. The normalisation constant $A$
is related to the internal energy density $U$ by, 
\beq
A=\frac{U\epsilon_{br,c}^{\gamma_1-2}}{[\frac{1}{\gamma_2-2}-\frac{1}
{\gamma_1-2}]}
\label{A_g}
\eeq 
where $\gamma_1 < 2$, and $\gamma_2 > 2$. The threshold energy for
pair production ($\gamma+\gamma\rightarrow e^{+}+e^{-}$) in the
comoving frame is $\sim 1$ MeV. Since the pair production
cross section is much larger than $p\gamma$ interaction cross
section, the upper cut-off energy in the photon spectrum which
produces high energy neutrinos should be limited by the threshold 
energy of pair production. Photon energy $1$ MeV in the comoving frame
corresponds to $300$ MeV and $10$ MeV in the source rest frame for 
HL and LL GRBs, with Lorentz factor of the order of $300$ and $10$,
respectively. For HL GRBs, $300$ MeV photons 
can produce neutrinos with energy $3000$ GeV in the source rest frame
in $p\gamma$ interactions through $\Delta$ resonance. For LL
GRBs, $10$ MeV photons can produce neutrinos with energy $100$ GeV
in the source rest frame.
The break energies of LL and HL GRBs are assumed to be of the order of
keV and MeV, respectively, in the source rest frame. Thus the upper
limits in photon energies are much higher than their break energies
for both LL and HL GRBs. The average comoving time 
is longer than the average synchrotron cooling time of protons
for HL GRBs, and for LL GRBs they are comparable. As we discuss
later in detail the average maximum neutrino energies are generally
constrained by synchrotron cooling of protons.  The corresponding 
lower limits in
photon energies for $p\gamma$ interactions are much lower than photon
spectral break energies for both HL and LL populations.  The
expression for parameter $A$ has been derived in Eq.(\ref{A_g}) using
the fact that the minimum photon energies are much lower than the
break energies and the maximum photon energies are much higher than
the break energies. The threshold energies for synchrotron self
absorption \cite{peer} of photons are of the order of eV for HL GRBs
and are much lower for LL GRBs. The lower limit in photon energies 
derived by the maximum proton energies (applying the synchrotron 
cooling constraint) is higher than the typical 
synchrotron self absorption energies for LL GRBs, and this is also
generally the case for HL GRBs.  If we take
$\gamma_1=1$, $\gamma_2=2$ and a photon energy range of 30KeV to 3MeV
in the source rest frame, then we get $A=\frac{U}{2\epsilon_{br,c}}$
as derived in \cite{wax1}.

In the comoving frame, the energy loss time ($t_\pi$) of a proton of energy $E_{p,c}$ due to pion production can be derived from
\beq
t_{\pi}^{-1}(E_{p,c})=-\frac{1}{E_{p,c}}\frac{dE_{p,c}}{dt}
\eeq
The variability time is of the order of milliseconds for HL GRBs.
For LL GRBs, observations show that their lightcurves are generally 
very smooth; for GRB 060218 the width of the broad peak was 431 sec
\cite{campana}. In $p\gamma$
interactions the maximum contribution to neutrino production comes
from the $\Delta$ resonance as discussed in \cite{wax1}. 
For $\gamma_1<2$ and $\gamma_2>2$, we obtain the fractional energy
loss of protons to pions as a function of proton energy $E_p$ [$=
E_p^{obs} (1+z)$] in the source rest frame.
\beq
f_{\pi}(E_p) =f_{0} \left\{\begin{array}{l@{\quad \quad}l}
\frac{1.34^{\gamma_2-1}}{\gamma_2+1}(\frac{E_{p}}{E_{pb}})^{\gamma_2-1}
& E_p<E_{pb}\\\frac{1.34^{\gamma_1-1}}{\gamma_1+1}
(\frac{E_{p}}{E_{pb}})^{\gamma_1-1} 
& E_{p}>E_{pb}\end{array} \right.
\label{fpi}
\eeq 
where, $f_{0}^{\rm HL}=\frac{0.9
L_{\gamma,51}}{\Gamma_{300}^4t_{v,-3}\epsilon_{br,MeV}}\frac{1}
{[\frac{1}{\gamma_2-2}-\frac{1}{\gamma_1-2}]}$ for HL GRBs and
$ f_{0}^{\rm LL}= \frac{0.729
L_{\gamma,47}}{\Gamma_{10}^4t_{v,2}\epsilon_{br,keV}}
\frac{1}{[\frac{1}{\gamma_2-2}-\frac{1}{\gamma_1-2}]}$ for LL GRBs.
Here $L_{\gamma,47}=L_\gamma/(10^{47}~{\rm ergs~s^{-1}})$ and 
$t_{v,2}=t_v/(10^{2}~{s})$.
For, $\gamma_1=1$ and $\gamma_2=2$ we get an expression of $f_{\pi}(E_p)$
for photon energy range 30keV to 3MeV similar to that derived in \cite{wax1}, 
\beq
f_{\pi}=0.2\frac{L_{\gamma,51}}{\Gamma^{4}_{300}t_{v,-3}\epsilon_{br,MeV}}
\left\{
 \begin{array}{l@{\quad \quad}l}
E_{p}/E_{pb}
& E_p<E_{pb}\\1. 
& E_{p}>E_{pb}\end{array} \right.
\label{wb}
\eeq 
where $\Gamma_{300}=\Gamma/300$, $\Gamma_{10}=\Gamma/10$ denote typical
Lorentz factors of HL and LL GRBs, respectively \cite{lith-zhang,dai-wang};
$\epsilon_{br,MeV}$ and $\epsilon_{br,keV}$ denote the break energy
$\epsilon_{br}$ in units of MeV and keV, respectively. The choice of a
low $\epsilon_{br}$ for LL GRBs is based on the fact that GRB 060218 has
$\epsilon_{br} \sim 5$ keV \cite{campana}, which well satisfy the
so-called Amati-relation \cite{amati,ggf}
\beq
\frac{\epsilon_{br}}{100keV}=(3.64\pm0.04)\Big(\frac{E_{\gamma}^{iso}}
{7.9\times10^{52}erg}\Big)^{0.51\pm0.01}
\eeq  
Another LL GRB 980425 apparently does not satisfy this correlation. 
However, Ghisellini et al. \cite{ghis1} argue that GRB 980425 might
not be an outlier. In particular, they suspect that {\em BeppoSAX}
only recorded a short duration of hard emission, and it might be
similar to GRB 060218 should it be detected by an instrument similar
to {\em Swift} that could record soft emission simultaneously. We
therefore assume the Amati-relation for all bursts in this paper
\cite{amati2}, but caution that the diffuse neutrino flux of the
LL-component would be overestimated (say, a factor of 2) if a fraction
(say, 50\%) of LL-GRBs do not satisfy the Amati relation. 
It is worth commenting that if most LL-GRBs satisfy the Amati-relation,
they are not detectable by BATSE, as is the case of GRB 060218
\cite{liang06b}, so that they do not contribute to the BATSE
population. The proton break energy corresponding to the break energy 
in the photon spectrum can be expressed as
\beq
E_{pb}^{\rm HL}=1.3\times10^{7}\Gamma_{300}^2(\epsilon_{br,MeV})^{-1}
{\rm GeV}. 
\eeq
for HL GRBs or
$E_{pb}^{\rm LL}=1.45\times10^{7}\Gamma_{10}^2
(\epsilon_{br,keV})^{-1}$ GeV for LL GRBs. In $p\gamma$ interactions both $\pi^{0}$ and $\pi^{+}$ can be
produced with equal probabilities. $\pi^{+}$ gets on the average
$20\%$ of the proton energy and if the final state leptons share the
pion energy equally then each neutrino carries $5\%$ of the initial
proton energy. The first break energy in the neutrino spectrum,
$E_{\nu b}$ is due to the break in the photon spectrum.
\beq
E_{\nu b}^{\rm HL}=6.5\times10^5\frac{\Gamma^2_{300}}
{\epsilon_{br,MeV}} {\rm GeV}
\label{Eb1}
\eeq
for HL GRBs or
$E_{\nu b}^{\rm LL}=7.23\times10^5\frac{\Gamma^2_{10}}
{\epsilon_{br,keV}}$ GeV for LL GRBs. 
The total energy to be emitted by neutrinos of energy
$E_{\nu}$ can be expressed as,
\beq
E_{\nu}^2\frac{dN_{\nu}(E_{\nu})}{dE_{\nu}}\approx
\frac{3f_{\pi}}{8}\frac{1}{\kappa}\frac{(1-\epsilon_e)}
{\epsilon_e}E_{\gamma}^{iso}
\label{spect}
\eeq
where $E_{\gamma}^{iso}$ is the total isotropic energy of the emitted
gamma-ray photons in the energy range of 1keV to 10MeV, which is
available from the observations\cite{frail}, $\epsilon_e \sim 0.3$ is 
the energy fraction carried by electrons, and $\kappa$ is a normalization
factor considering the differential neutrino spectrum. The relativistic 
electrons produce the photons by synchrotron radiation and inverse 
Compton scattering of low energy photons, so four orders of magnitude 
in photon energy corresponds to two orders of magnitude in the energy 
of the radiating charged leptons. Photon spectral index $\gamma_2=2.25$ 
corresponds to a spectral index of $-2.5$ of the relativistic electron 
spectrum. This correspond to $\kappa=1.8$ assuming photon fluence is 
proportional to neutrino luminosity. If the electron spectral index is 
$-2$ one would have $\kappa = \ln(100) \sim 4.6$. 
The internal shocks of GRBs are typically collisionless and Coulomb 
interaction is not important between electrons and protons. It is still
an open question regarding how some energies of Fermi accelerated 
protons are transferred to electrons and whether electrons and protons
would have a same spectral index. In principle, the spectral indices of
electrons and protons in the shocked region could be different.
In our calculations, the electron spectral index is taken as $-2.5$ 
(based on the observed photon spectrum), while the proton spectral index 
is still assumed to be $\sim (-2)$ as has been adopted by all the 
previous neutrino calculations.

 The break (Eq.(\ref{Eb1}))
due to the break energy in the photon spectrum is caused by the break
in $f_{\pi}$ (Eq.(\ref{fpi})).  If one takes into account muon and
pion cooling, a second break appears in the neutrino spectrum
\beq
 E_{\nu}^2\frac{dN_{\nu}(E_{\nu})}{dE_{\nu}}\approx\frac{3f_{\pi}}
{14.4}\frac{(1-\epsilon_e)}{\epsilon_e}E_{\gamma}^{iso}
 \left\{\begin{array}{l@{\quad \quad}l} 1 & E_{\nu}<E_{\nu}^{s}\\
 (\frac{E_{\nu}}{E_{\nu}^s})^{-2}& E_{\nu}>E_{\nu}^{s} \end{array}
 \right.
\eeq 
Pion cooling energy is ten times
higher than the muon cooling energy. We would be overestimating the
total neutrino flux if we use the pion cooling energy to derive
the second break energy in the neutrino spectrum.
The muon cooling break $E_{\nu}^s$ can be expressed as a function of
GRB parameters. 
\beq
E_{\nu}^{s,{\rm HL}}=2.56\times10^{6}\epsilon_e^{1/2}\epsilon_B^{-1/2}
L_{\gamma,51}^{-1/2}\Gamma_{300}^4t_{v,-3}{\rm GeV}
\eeq
for HL GRBs or $E_{\nu}^{s,{\rm
LL}}=3.16\times10^{7}\epsilon_e^{1/2}\epsilon_B^{-1/2}
L_{\gamma,47}^{-1/2}\Gamma_{10}^4t_{v,2}$GeV for LL GRBs. Here the
parameter $\epsilon_B$ denotes the fraction of internal energy carried
by the magnetic fields. We can see that if $\epsilon_e$ and
$\epsilon_B$ are comparable their contributions are cancelled
out. They are assumed to be equal to $0.3$ in our calculations. 
Neutrino oscillation only redistributes the number of different
flavors of neutrinos. The total number of neutrinos remains unchanged. 

The maximum proton energy achievable in HL and LL GRBs can be derived
by equating the acceleration time scale with the minimum of the 
dynamic time scale and synchrotron cooling time scale of protons
\cite{lect}. 
In the comoving frame, the synchrotron cooling time can be expressed 
in terms of proton energy in the source rest frame as, 
\beq
t_{sync}=\Gamma \frac{6\pi m_{p}^4 c^3}{\sigma_{t} m_e^2 E_p B_{c}^2}~,
\label{syntime}
\eeq
where $m_p$ and $m_e$ are proton and electron masses respectively,
$\sigma_{t}$ is the Thompson scattering cross section and $B_c$ is the
comoving magnetic field. The comoving magnetic field strength
can be expressed as a function of radial distance, luminosity, 
Lorentz factor the unknown shock strength (which depends on the
relative Lorentz factor of the two colliding shells), and the unknown
shock equaiartition parameters \cite{bp}. For $\epsilon_B=0.3$ and
reasonable shock strength, one could estimate
$B_{c}^{HL}\simeq(2.5\times10^{5}G)L_{51}^{1/2}r_{13}^{-1}$
with $r^{HL}_{13}=0.27\Gamma^2_{300}t_{v,-3}$ for HL and
$B_{c}^{LL}\simeq (4.42\times10^{4}G) L_{47}^{1/2} r_{13}^{-1}$,
$r^{LL}_{13}= 30\Gamma^2_{10}t_{v,2}$ for LL GRBs, $r_{13}=r/10^{13}$ is the dissipation radius where internal shocks are formed. The acceleration
time scale in the comoving frame in terms of proton energy in the
source rest frame is,
\beq
t_a=\frac{E_p}{\Gamma B_{c} e c}
\eeq
The dynamical comoving time scale is
\beq
t_{dyn}=\frac{r}{\Gamma c}~.
\eeq
In our calculations, we use $t_a = {\rm min} (t_{sync}, t_{dyn})$ to
calculate the maximum proton energy in the shock. For typical parameters,
one usually has $t_{sync} < t_{dyn}$, one can then derive the expression 
of the maximum proton energy using the condition $t_{sync}=t_a$, which
gives
\beq
E_{p,max}^2=\Gamma^2 \frac{6\pi m_p^4 c^4e}{\sigma_t m_e^2 B_{c}}~,
\eeq 
which gives $E_{p,max}^{HL}\approx 6\times10^{10}\frac{\Gamma_{300}}
{\sqrt{B^{HL}_{c,4}}}$ GeV for HL GRBs and $E_{p,max}^{LL}\approx
2\times10^{10}\frac{\Gamma_{10}}{\sqrt{B^{LL}_{c,2}}}$ GeV for LL GRBs.  
After averaging over the parameters, including log normal distributions 
in Lorentz factor and variability time, the broken power-law
distribution of the luminosity distribution, and the redshift
distribution of GRBs (see below for detail), we obtain the average 
values of maximum proton energies for LL and HL GRBs.

The neutrino spectrum emitted from the source is to be corrected for the
redshift of the GRB to derive the total spectrum to be observed on
earth. The total isotropic energy can be expressed as a product of the
isotropic luminosity ($L_{\gamma}$) and the duration of the burst
($T_d$).  The observed neutrino spectrum on earth
$\frac{dN_{\nu}^{ob}(E_{\nu}^{ob})}{dE_{\nu}^{ob}}$ is a function of
$\Gamma$, $L_{\gamma}$, $T_d$, $z$, $t_v$ and $E_{\nu}^{ob}$
\beq
\frac{dN_{\nu}^{ob}(E_{\nu}^{ob})}{dE_{\nu}^{ob}}=\frac{dN_{\nu}(E_{\nu})}
{dE_{\nu}}\frac{1}{4\pi d^2(z)}(1+z) 
\eeq
where $r(z)$ is the comoving radial coordinate distance of the
GRB. For a spatially flat universe with $\Omega_{\Lambda}+\Omega_m=1$,
the comoving distance can be expressed as
\beq
d(z)=\int_0^z\frac{c}{H_0}\frac{dz^{\prime}}
{\sqrt{\Omega_{\Lambda}+\Omega_{m}(1+z^{\prime})^3}}
\eeq 
In our calculations we use $\Omega_{\Lambda}=0.73$, $\Omega_{m}=0.27$
and the Hubble constant $H_0=71 ~{\rm km ~sec^{-1}~ Mpc^{-1}}$ from
\cite{ben}. When calculating the diffuse neutrino background from
GRBs we assume 
that GRBs follow the star formation rate\footnote{If however GRBs
do not trace the global star-forming rate, our normalization factor
needs to be modified.}. The diffuse neutrino
flux on earth from GRBs distributed up to a redshift of $z_{max}$ can
be expressed as,
\beq
M_{\nu}^{ob}(L_{\gamma},\Gamma,T_d,t_v,E_{\nu}^{ob})=
\int_0^{z_{max}}\frac{dN_{\nu}^{ob}(E_{\nu}^{ob})}
{dE_{\nu}^{ob}}\frac{R_{GRB}(z)}{1+z}dV(z) 
\eeq
where, the star formation rate is given by \cite{prc}
\beq
R_{GRB}(z)=23\rho_0\frac{e^{3.4z}}{e^{3.4z}+22}~.
\eeq 
The local rates $\rho_0$, of the LL and HL populations have been
assumed to be about $550 {\rm Gpc^{-3}yr^{-1}}$ and $1.1 {\rm
Gpc^{-3}yr^{-1}}$, respectively \cite{liang}. We obtain the neutrino
fluxes from LL and HL populations distributed up to a distance of
$z_{max}=5$. The neutrino fluxes have been averaged with luminosity, 
Lorentz factor, variability time and burst duration distribution 
functions to take into account their variations in the population
(the variations in luminosity and duration also reflect variation
in isotropic energy)
\bea
\phi_{\nu}(E_{\nu}^{ob})=\int_{L_{\gamma,1}}^{L_{\gamma,2}}
\int_{\Gamma_{1}}^{\Gamma_{2}}\int_{T_{d1}}^{T_{d2}} \int_{t_{v1}}^{t_{v2}}
M_{\nu}^{ob}(L_{\gamma},\Gamma,T_d,t_v,E_{\nu}^{ob}) 
S(L_{\gamma})G(\Gamma)P(T_d) Q(t_v)dL_{\gamma} d\Gamma dT_{d}dt_{v}.
\label{average}
\eea
Here $S(L_\gamma)$ is the broken power law luminosity function
(eqn(2)), with $L_b^{\rm HL}=10^{51}~{\rm erg ~s^{-1}}$,
$\alpha_1^{\rm HL} \sim1.15$, $\alpha_2^{\rm HL}\sim2.5$ for HL and
$L_b^{\rm LL}=10^{47} ~{\rm erg~s^{-1}}$, $\alpha_1^{\rm LL}\sim0.6$
and $\alpha_2^{\rm LL} \sim4.5$ for LL GRBs (following \cite{liang}).  
For the HL population
the lower and upper limits in luminosities are $L_{\gamma,1}^{\rm
HL}=10^{50}~{\rm erg~s^{-1}}$ and $L_{\gamma,2}^{\rm HL}=10^{52}~ 
{\rm erg~s^{-1}}$, respectively, and the corresponding 
values for the LL population are $10^{46}~{\rm erg~s^{-1}}$ and
$5\times10^{49}{\rm erg~s^{-1}}$, respectively. 
The Lorentz factor distribution
function $G(\Gamma)$ is asssumed to be log normal. For HL GRBs, we
assume the mean of the distribution at $\log ({\Gamma_m})=2.6$, which
correspond to $ \Gamma_m\approx 398$, with a standard deviation of
0.3. For LL GRBs, we assume the log normal distribution of Lorentz
factors is peaked at $\log({\Gamma_m})=1.1$ \cite{dai-wang} with a
standard deviation 0.1. We allow the values of Lorentz factors to
vary from 100 to 1000 for the HL GRBs, and from 5 to 50 for the LL
population. The durations of the bursts are also assumed to follow log
normal distributions $P(T_d)$, with the mean at $\log(T_{d,m})=1.47, 
3.5$; which
corresponds to about 30sec and 3000sec, for the HL and LL populations,
respectively, with standard deviations 0.5 and 0.1. The
scattering in Amati's relation \cite{ggf} does not affect our results
qualitatively. Since the photon index $\gamma_2$ does not differ
significantly among bursts \cite{saka}, we take $\gamma_2=2.25$
universally for both the HL and LL populations.
We also consider variation in variability time of GRBs. Their
distributions are assumed to be log normal, represented by $Q(t_v)$ in
eqn.(\ref{average}) with the mean at $\log(t_{v,m})=-1.52, 2.6$ and
standard deviations 0.3, 0.1 for HL, LL GRBs respectively. The maximum
energies of neutrinos have also been averaged with the Lorentz factor,
luminosity and variability time distribution functions as discussed
above.
\bea
E^{av}_{\nu,max}=\int_{L_{\gamma,1}}^{L_{\gamma,2}}
\int_{\Gamma_{1}}^{\Gamma_{2}}\int_{t_{v1}}^{t_{v2}}
E_{\nu,max}(L_{\gamma},\Gamma,t_v) 
S(L_{\gamma})G(\Gamma)Q(t_v)dL_{\gamma}d\Gamma dt_{v}.
\eea
Also, the redshift correction on the observed neutrino energy has been 
averaged over the entire volume between $z=0$ to $z=5$.
\beq
E^{av,ob}_{\nu,max}=E^{av}_{\nu,max}\frac{\int_{0}^{5} 
R_{GRB}(z)dV(z)/(1+z)}{\int_{0}^{5} R_{GRB}(z)dV(z)}
\eeq
  
\section{Results and Discussions} 
The neutrino spectra from HL and LL populations are plotted in units
of ${\rm GeV~ cm^{-2}~ s^{-1}~ sr^{-1}}$ in Fig.1. for $\gamma_1=1$,
1.1, 1.3 and 1.5 and $\gamma_2=2.25$.  We have assumed
$\epsilon_e=0.3$ and $\epsilon_B=0.3$ in our calculations. The long
dashed (green) lines represent neutrino fluxes from the LL population
and the solid (red) lines are for the HL population. The
Waxman-Bahcall \cite{limit1} limit with $z$ evolution (including all
three flavors) has been shown by short dashed (blue) lines, and the
dotted (pink) lines represent the sensitivity of the IceCube detector
\cite{ice2} after three years of operation. The current limit from
AMANDA-II data taken between 2000 and 2003 \cite{amanda} is also shown
by triple dotted line above the WB upper bound. The HL component has
an average maximum energy of $6.3\times10^9$ GeV and for LL population
it is $1.5\times10^9$ GeV. Above about $10^8$ GeV the contribution of
LL population to the diffuse neutrino background is more than the HL
population. The LL component continues to below $100$ GeV but the HL
component has a lower cut-off energy of about $1000$ GeV. Between
$1000$ GeV and $4\times 10^{4}$ GeV their fluxes are equal. We have
also plotted the GZK neutrino flux from \cite{stecker} in thin dashed
lines for a comparison.  Comparing with the Icecube sensitivity, we
can see that the GZK neutrinos are generally not detectable by
Icecube, while the contribution of the LL-component would be
detectable if $\gamma_1\geq1.1$. In this case we expect
30$km^{-2}sr^{-1}$ neutrino events per year or more in IceCube at an
energy of $10^7$GeV. If we compare our derived neutrino spectra from
HL GRBs with some of the previous results
\cite{raz2} the width of the intermediate region between the two break
energies is narrower. This is because we have used the muon cooling
energy instead of the pion cooling energy in determining the second
break energy of the spectra to avoid overestimation of neutrino
fluxes. However, in this way we slightly underestimate the $\nu_{\mu}$
fluxes as they are produced by $\pi^{+}$ decay unlike $\bar\nu_{\mu}$
and $\nu_e$.  We also find that our results are very sensitive to
variations in the mean value of Lorentz factor. As the break energies
in the neutrino spectra depend on $\Gamma^2$ and $\Gamma^4$ a slight
deviation in the value of $\Gamma$ changes the neutrino fluxes
significantly. In the future it might be possible to set limits on the
values of GRB Lorentz factors if the detection or upper limit of 
high energy neutrino background from GRBs is set up. Overall, our
results show that the previously neglected LL population of GRBs
significantly contribute to the diffuse neutrino background as their
event rate is very high. This conclusion would be strengthened as {\em
Swift} detects more LL GRBs in the years to come.

Finally we notice that after we submitted the first version of this
paper to astro-ph, similar results were independently reported by
Murase et al.\cite{murase}.

\begin{figure*}[t]
\begin{center}
\centerline{
\epsfxsize=15.cm\epsfysize=10.cm
\epsfbox{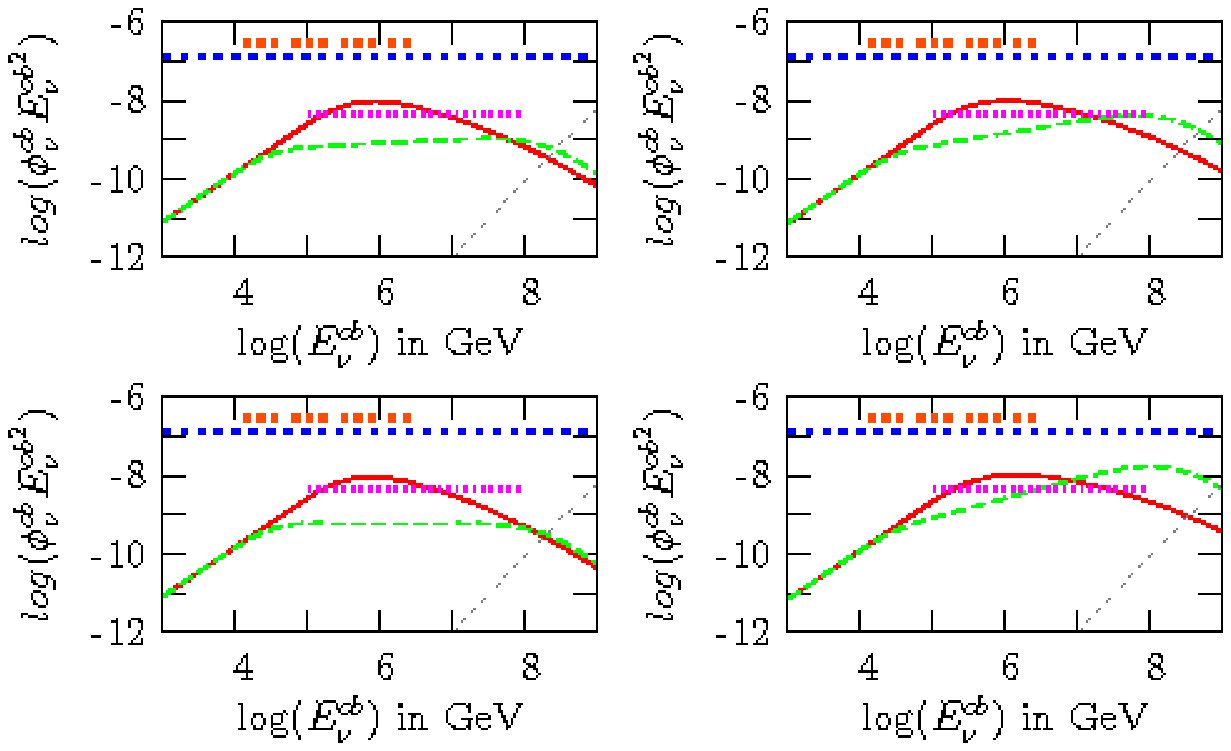} }
\end{center}
\caption{The diffuse neutrino fluxes expected on earth from low and high
luminosity populations of GRBs distributed up to a redshift of 5,
being plotted in units of ${\rm GeV ~cm^{-2}~ s^{-1} sr^{-1}}$ against
the observed neutrino energy. The four panels are for photon spectral 
indices $\gamma_1=1, 1.1, 1.3, 1.5$ (clockwise from left bottom), 
and $\gamma_2=2.25$ are adopted for all of them. The
Waxman-Bahcall limit with $z$ evolution \cite{limit1} has been shown
with short dashed (blue) line. The double dotted (pink) line
represents IceCube experiment's sensitivity \cite{ice2} after 3 years
of operation and the current limit from AMANDA-II data (2000-2003)
\cite{amanda} triple dotted (orange) line is just above the
Waxman-Bahcall limit. These limits include all the three flavors of
neutrinos. GZK neutrino flux has been plotted in thin dashed line 
from \cite{stecker} for comparison.}
\label{fig:Nnu}
\end{figure*}

We thank the anonymous referee for constructive reports that help
to improve the paper.  This work was supported by NASA under grants 
NNG05GB67G and NNG06GH62G.


\begin{thebibliography}{99}
\bibitem{amati2} Different populations of GRBs (e.g. HL and LL) could
well share a same emission mechanism that the Amati-relation
obeys. Therefore that both LL and HL GRBs satisfy the Amati-relation
should not be regarded an argument against they belonging to two
distinct population groups.
\end{thebibliography}
\end{document}